\documentclass[aps,prl,reprint,groupedaddress,showpacs]{revtex4-1}

\usepackage{latexsym,amssymb,amsfonts,mathrsfs,color,graphicx,amsmath}

\newcommand{\Per}{\text{Pe}_r}
\newcommand{\Ueff}{V_{\text{eff}}}

\begin{document}

\title{Detention times of microswimmers close to surfaces: \\
Influence of hydrodynamic interactions and noise}
\author{Konstantin Schaar$^{1,2,3}$}
\author{Andreas Z\"{o}ttl$^{1}$}
\author{Holger Stark$^{1}$}
\date{\today}

\affiliation{$^{1}$Institut f\"{u}r Theoretische Physik, Technische Universit\"{a}t Berlin, Hardenbergstrasse 36, 10623 Berlin, Germany}
\affiliation{$^{2}$\mbox{Institut f\"{u}r Theoretische Biologie, Humboldt Universit\"{a}t Berlin, Invalidenstrasse 43, 10115 Berlin, Germany}}
\affiliation{$^{3}$Robert Koch-Institut, Seestrasse 10, 13353 Berlin, Germany}

\begin{abstract}
After colliding with a surface, microswimmers reside there during the detention time. They accumulate and may form complex structures such as biofilms. We introduce a general framework to calculate the distribution of detention times using the method of first-passage times and study how rotational noise and hydrodynamic interactions influence the escape from a surface. We compare generic swimmer models to the simple active Brownian particle. While the respective detention times of source dipoles are smaller, the ones of pullers are larger by up to several orders of magnitude, and pushers show both trends. We apply our results to the more realistic squirmer model, for which we use lubrication theory, and validate them by simulations with multi-particle collision dynamics.
\end{abstract}

\maketitle

Biological microswimmers such as bacteria are omnipresent in our everyday life. At the micron scale their locomotion in aqueous environment is determined by low-Reynolds-number hydrodynamics and influenced by thermal and intrinsic biological noise \cite{Berg93,Lauga09}. In real environments such as the human body \cite{Bray00} or the ocean \cite{Callow02,Rosenberg07} microorganisms swim in the presence of soft or solid boundaries where they may form complex aggregates such as biofilms \cite{Watnick00}. This letter develops a general approach for investigating the fundamental and biologically relevant question how long a swimming microorganism resides at bounding surfaces by accounting for both hydrodynamic swimmer-wall interactions and noise.

To develop an understanding for the accumulation and the dynamics of microorganisms near walls, several important aspects have been investigated recently: swimmer-wall hydrodynamic  interactions \cite{Lauga06,Berke08,Drescher09,Drescher11}, thermal and intrinsic noise \cite{Li09,Drescher11}, cilia- and flagella-wall interactions \cite{Kantsler13}, bacterial tumbling \cite{Molaei14}, and buoyancy \cite{Jung14}. Whether stochastic motion or swimmer-wall hydrodynamic interactions determine the reorientation of microswimmers at a surface  and how they both influence the bacterial distribution between parallel plates has been discussed controversially \cite{Berke08,Li09,Drescher11}. Hydrodynamic interactions trap bacteria at surfaces \cite{Berke08,Spagnolie12}, force them to swim in circles \cite{DiLuzio05}, or even suppress bacterial tumbling \cite{Molaei14}. However, non-tumbling bacteria\ \cite{Li09,Drescher11} or elongated artificial microswimmers\ \cite{Takagi14} use rotational noise to escape from surfaces.

Artificial microswimmers such as active Janus particles or squirmers, which are driven by a surface velocity field, have been studied in front of a no-slip wall both in experiments \cite{Volpe11,Kreuter13} and by theoretical models. The latter either include hydrodynamic interactions \cite{Crowdy10,Spagnolie12,Crowdy13,Ishimoto13,Uspal14,Li14} or only consider active Brownian particles \cite{vanTeeffelen08,Volpe11,Enculescu11,Elgeti13,Lee13}.

\begin{figure}[b]
\includegraphics[width=0.9\columnwidth]{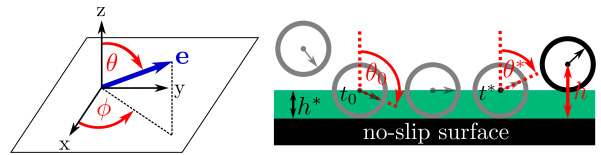}
\caption{Definition of coordinate system and sketch of a typical trajectory for a spherical microswimmer approaching a plane no-slip surface ($h=h^{*}$) at time $t_{0}$ and leaving the surface at $t^{*}$. The detention time at the surface is $t^{*}-t_{0}$.}
\label{Fig:SketchSequenz}
\end{figure}

An important prerequisite for the observed accumulation near walls are the relatively large times microswimmers reside at a surface before leaving it \cite{Volpe11,Takagi14}. In this article we call these swimmer-wall contact times detention times and 
 calculate their distributions near a plane no-slip surface based on the method of first-passage times \cite{Honerkamp90}. For generic microswimmers we demonstrate that hydrodynamic interactions, relative to pure rotational noise, can either increase the mean detention time by several orders of magnitude or also decrease it.

At low Reynolds number the motion of an axisymmetric microswimmer with orientation $\mathbf{e}$ in the presence of bounding surfaces is governed by the Langevin equations
\begin{align}
\begin{split}
\dot{\mathbf{r}}&=\mathbf{v}_{\text{A}}+\mathbf{v}_{\text{HI}}+\mathbf{v}_{\text{N}}+...,\\
\dot{\mathbf{e}}&=\boldsymbol{\Omega}\times\mathbf{e}~~\text{with}~~
\boldsymbol{\Omega}=\boldsymbol{\Omega}_{\text{HI}}+\boldsymbol{\Omega}_{\text{N}}+...,
\label{Eq:1}
\end{split}
\end{align}
which account for the stochastic dynamics of position $\mathbf{r}$ and orientation $\mathbf{e}$. Here we only consider the influence of the activity of the swimmer ($\mathbf{v}_{\text{A}}=U\mathbf{e}$ with bulk swimming velocity $U$), hydrodynamic interactions with the surface (HI), and noise (N). However, our approach can in principle be used for any dynamics which is of the form of Eqs.\ (\ref{Eq:1}) and also include, e.g., steric or electrostatic interactions as well as external fluid flow.

We consider a spherical microswimmer, moving on a smooth trajectory, which reaches the wall at time $t_0$ with an angle $\theta_0$ against the surface normal (see Fig.\ \ref{Fig:SketchSequenz} and a typical trajectory in the Supplemental Material \cite{supp}). This occurs at P\'eclet number $ \mathrm{Pe} = U R / D_t \gg 1$, where $R$ is the radius and $D_t$ the translational diffusion coefficient of the swimmer. Typical values are $\text{Pe} \gtrsim 10^2$ for bacteria, $\text{Pe}\gtrsim 10^3$ for sperm cells and $\text{Pe}\gtrsim 10^4$ for \textit{Chlamydomonas}. The swimmer stays at a height  $h \approx R$, so we neglect translational motion in the following \cite{noteTrans}. The swimming direction $\mathbf{e}$ diffuses on the unit sphere but also drifts with angular velocity $\boldsymbol{\Omega}_{\text{HI}} = \Omega_{\text{HI}} \mathbf{e}_{\phi}$. Once the swimming direction has reached the escape angle $\theta^{*}$, to be defined below for each swimmer type, the microswimmer leaves the surface at time $t^{*}$. This stochastic process is described by the Smoluchowski equation $\partial_{t} P = {\cal L} P = (- {\cal R} \cdot \boldsymbol{\Omega}_{\text{HI}} + D_{r} {\cal R}^2) P$, where  ${\cal R} =  \mathbf{e} \times \boldsymbol{\nabla}_{\mathbf{e}} $ is the rotation operator and $D_r$ the rotational  diffusion constant\ \cite{Enculescu11,Wolff13}.

Rotational diffusion along the azimuthal angle $\phi$ does not influence the escape from the surface and it is sufficient to consider the conditional probability $p(\theta, t^{\ast} | \theta_0,t_0)= \int_{0}^{2\pi} \mathrm{d}\phi_0 \int_{0}^{2\pi} \mathrm{d}\phi P(\theta, \phi, t^{\ast} | \theta_0,\phi_0,t_0)$. To calculate the distribution of detention times at the surface, we use the Fokker-Planck approach of first-passage problems \cite{Honerkamp90}. The integrated probability $g( \theta^{\ast}, t | \theta_{0} ) = \int_{\theta^{\ast}}^{\pi} p(\theta, t^{\ast} | \theta_0,t_0) \sin \theta \mathrm{d} \theta$ for finding the swimming direction in the angular interval $[\theta^{*}, \pi]$ at time $t = t^{\ast} - t_0$ obeys the adjoint Smoluchowski equation (see \cite{supp})
\begin{align}
\partial_{t} g( \theta^{\ast}, t | \theta_{0} )  = {\cal L}^{+}(\theta_0) g( \theta^{\ast}, t | \theta_{0})
\label{Ref-ADJFPE1D},
\end{align}
with ${\cal L}^{+}(\theta_0) = \Omega (\theta_0) \partial_{\theta_{0}} + D_{r}  \partial^{2}_{\theta_{0}} $, where $\Omega(\theta_0) =  \Omega_{\text{HI}} (\theta_0) + D_r\cot \theta_0$ is an effective angular drift velocity. To solve it, one uses at $\theta_0 = \pi$ reflective [$  \left. \partial_{\theta_{0}} g( \theta^{\ast}, t | \theta_{0}) \right|_{\pi} =0$] and at $\theta_0 = \theta^{\ast}$ absorbing  [$ g( \theta^{\ast}, t | \theta^{\ast}) =0$] boundary conditions. Now, $-\partial_{t}g(\theta^{*},t|\theta_{0}) \mathrm{d}t$ is the probability to leave the surface with escape angle $\theta^{\ast}$ at time $t$ in the time interval $\mathrm{d}t$, so 
\begin{align}
f(\theta^{*},t|\theta_{0})=-\partial_{t}g(\theta^{*},t|\theta_{0})
\label{Relation-f-and-g}
\end{align}
denotes the distribution of detention times $t=t^{\ast} - t_0$ for being trapped at the surface (DTD).

To investigate how hydrodynamic interactions compared to pure rotational noise influence the detention time, we calculate the DTD $f(\theta^{*},t|\theta_{0})$ for several model microswimmers by numerically solving Eq.~(\ref{Ref-ADJFPE1D}) and using Eq.~(\ref{Relation-f-and-g}). From here on, we always rescale time by the ballistic time scale $\tau_{s}=R/U$ and introduce the persistence number $\text{Pe}_{r} = (2  D_{r} \tau_s)^{-1} $. Since $(2D_{r})^{-1}$ is the orientational correlation time, $\text{Pe}_{r} \gg 1$ means directed swimming \cite{Li09,Taktikos12}. Typical values are $\text{Pe}_r \gtrsim  100$ for sperm cells \cite{Friedrich09} and non-tumbling \textit{E.~coli} \cite{Drescher11}, or $\text{Pe}_r \approx 25$ for \textit{Chlamydomonas} \cite{Drescher11}.

\begin{figure}[t]
\includegraphics[width=0.95\columnwidth]{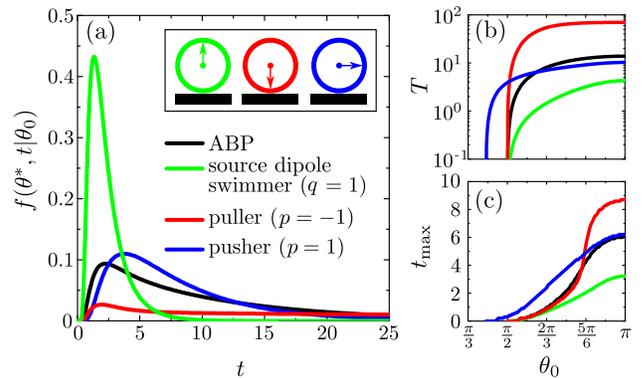}
\caption{(a) DTD for ABP and source- and force-dipole swimmer with $\text{Pe}_{r}=10$ and an initial angle $\theta_{0}=3\pi/4$.  
 (b) Mean detention time $T$ versus initial angle $\theta_{0}$. (c) Most likely detention time $t_{\text{max}}$ (maximum of $f$).}
\label{Fig:Fig2}
\end{figure}

First, we consider a spherical active Brownian particle (ABP) with $\Omega_{\text{HI}}=0$ near a surface \cite{Enculescu11,Elgeti13}. The escape angle is simply $\theta^{*} = \pi/2$. From the known propagator of free rotational diffusion \cite{Berne00}, one can determine $g( \theta^{\ast}, t | \theta_{0} )$ and ultimately the DTD becomes
\begin{align}
\begin{split}
f\Big(\frac{\pi}{2},t |\theta_{0}\Big) &= \frac{\pi}{2\text{Pe}_r}  \sum_{l=1, \, \mathrm{odd} \, l}^{\infty}
 (-1)^{\frac{l+1}{2}}
e^{-l(l+1) t / (2\mathrm{Pe}_r)}  \\
& \times  \frac{l(2l+1)}{2^{l-1}} 
\left( \begin{array}{c}   l -1 \\  \frac{l-1}{2} \end{array} \right) P_l(\cos \theta_0) \, ,
\label{f-neutral-swimmer}
\end{split}
\end{align}
where $P_l(\cos \theta_0)$ are Legendre polynomials. The DTD is plotted in  Fig.~\ref{Fig:Fig2}(a) for $\theta_0=3\pi/4$ and $\text{Pe}_r=10$. The mean detention time $T = \int_0^\infty t f(\theta^{*},t|\theta_0)\mathrm{d}t $ of the ABP at the surface is calculated following Ref. \cite{Honerkamp90},
\begin{align}
T^{\text{ABP}} = 2\text{Pe}_{r} \ln(1-\cos\theta_0) \, .
\end{align}
We plot $T^{\text{ABP}}$ versus $\theta_0$ in Fig.~\ref{Fig:Fig2}(b). Note that the most likely detention time $t_{\text{max}}$ [see Fig.~\ref{Fig:Fig2}(c)] is much smaller compared to $T^{\text{ABP}}$ due to the slow decay of $f(\theta^{*},t|\theta_0)$.

Second, we consider microswimmers which generate either a force-dipole flow field of strength $p$ or a source dipole field of strength $q>0$ in the surrounding fluid \cite{Lauga09}. Examples for the first case are pushers ($p>0$) such as bacteria or pullers ($p<0$) such as the biflagellated algae \textit{Chlamydomonas}. Source dipoles are realized by active droplets \cite{Thutupalli11} or \textit{Paramecia} \cite{Ishikawa06b}. Each flow field is described by a flow singularity located in the center of the swimmer. For simplicity, we assume that the description by singularities is still valid close to the wall (see also the discussion in  \cite{Drescher11,Spagnolie12}). Their flow fields interact hydrodynamically with the surface and thereby generate wall-induced angular velocities $\Omega_{\text{HI}}$ of the microswimmers. At the wall ($h=R$) they read $\Omega_{\text{HI}}=3 p \, \sin \theta \cos \theta /8$ for the force dipole and $\Omega_{\text{HI}}=-3 q \, \sin\theta/8$ for the source dipole, respectively \cite{Berke08,Spagnolie12,timedependent}. The stable orientations $\theta_s$ of our swimmer types at the wall in the absence of noise are sketched in the inset of Fig.~\ref{Fig:Fig2}(a). They are calculated from $\Omega_{\text{HI}}(\theta_s)=0$ and $\partial \Omega_{\text{HI}} (\theta) / \partial \theta |_{\theta= \theta_s}<0 $. 

Hydrodynamic interactions of the source dipole ($q>0$) always rotate the swimmer away from the surface until it leaves the surface at $\theta^{*} = \pi/2$. Hence, the width of the DTD is much narrower compared to the ABP [see Fig.~\ref{Fig:Fig2}(a)]. The mean detention time $T$ plotted in Fig.~\ref{Fig:Fig2}(b) is much smaller compared to $T^{\text{ABP}}$ for all incoming angles $\theta_0$ due to $\Omega_{\text{HI}} \propto - q$ and the most likely detention time $t_{\text{max}}$ is comparable to $T$ [see Fig.~\ref{Fig:Fig2}(c)]. 

The puller ($ p <0$) is rotated towards the surface by hydrodynamic interactions if $\theta > \pi/2$ and can only escape if angular noise drives it to $\theta < \theta^{\ast} = \pi/2$. As a consequence, the  DTD only has a weakly pronounced maximum and decays very slowly [see Fig.~\ref{Fig:Fig2}(a)]. Therefore, at $\text{Pe}_r = 10$ the mean detention time of the puller is by an order of magnitude larger than for the ABP. We note that for biological swimmers direct flagella-wall interactions can significantly influence the reorientation at the wall. For the puller algae \textit{Chlamydomonas} $\Omega_{\text{steric}}>0$, which rotates the cell away from the surface \cite{Kantsler13} and strongly decreases the detention times compared to ABPs \cite{supp}.

The situation of the pusher ($ p > 0$) is more complex. Due to hydrodynamic interactions it has a stable orientation parallel to the wall [$\theta_s=\pi/2$, see inset of Fig.~\ref{Fig:Fig2}(a)]. Since, in addition, the wall-induced velocity $\mathbf{v}_{\text{HI}}(\theta_s)$ pushes it towards the wall, a noiseless pusher always  swims at the wall \cite{Berke08} and $T \rightarrow \infty$. In the presence of noise the swimmer orientation fluctuates about its stable direction. The pusher stays trapped until the escape angle $\theta^{*} < \pi/2$ is reached, where the total swimmer velocity starts to point away from the wall. Thus, the escape angle is determined by the condition $[\mathbf{v}_{\text{A}}(\theta^{*}) + \mathbf{v}_{\text{HI}}(\theta^{*}) ] \cdot \mathbf{e}_{z} = 0$, which gives $\theta^{*}=\arccos[(-4+\sqrt{16+27 p^2})/(9 p)]$ \cite{Berke08,Drescher11}.

Hydrodynamic interactions of the pusher with the surface can either enhance or reduce the detention time compared to  an ABP. On the one hand, increasing $ p  \propto \Omega_{\text{HI}}$ from zero reduces the time to reach the stable  orientation  and thus the time to get closer to the escape angle $\theta^{*} < \pi/2$. This can reduce the mean detention time compared to ABPs for small $ p $ as illustrated in Fig.~\ref{Fig:Fig2}(b). On the other hand, increasing $p$ further traps the orientation more strongly at $\theta_s=\pi/2$  and also pushes $\theta^{*}$ more and more away from $\theta_s$. Since rotational diffusion has to compensate for both effects, the detention time in\-crea\-ses.

\begin{figure}[t]
\includegraphics[width=0.94\columnwidth]{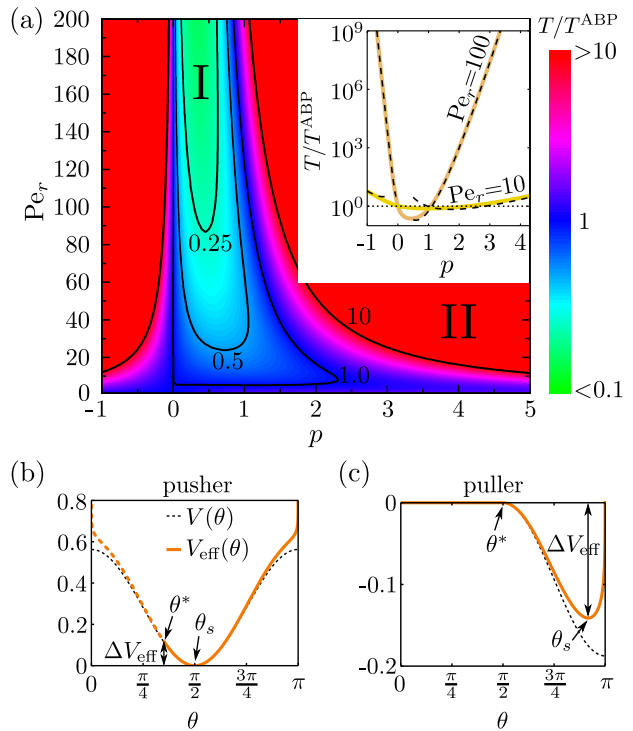}
\caption{
(a) Mean detention time $T/T^{\text{ABP}}$ for the force-dipole swimmer plotted versus $ p$ and $\text{Pe}_{r}$ for $\theta_{0}=3 \pi / 4$. Within region I, $T/T^{\text{ABP}}  < 1$, while in region II, $T/T^{\text{ABP}} \gg 1$. Inset: $T(p)/T^{\text{ABP}}$ for two values of $\text{Pe}_{r}$ and compared to Eqs.\ (\ref{Eq:TPush}) and (\ref{Eq:TPull}) (dashed lines). (b,c)  Effective angular potentials $V_{\mathrm{eff}}(\theta)$ and deterministic potentials $V(\theta)$ ($\text{Pe}_r\rightarrow\infty$) for a pusher (b) ($p=3$) and a puller (c) ($p=-1)$ at $\text{Pe}_r=20$.}
\label{Fig:Fig3}
\end{figure}

Figure \ref{Fig:Fig3}(a) gives an overview of the force-dipole swimmer by plotting $T / T^{\text{ABP}}$ in a color code versus $\text{Pe}_r$ and $p$. For negative $p$ the strong increase of $T$ beyond $T^{\text{ABP}}$ with increasing $|p |$ is visible and also documented in the inset for two values of $\text{Pe}_r$. For small positive $ p $ and for $\text{Pe}_r  \gtrsim 5 $ a clear minimum of $T$ develops as just discussed (see also the inset). In particular, in region I one finds $T  < T^{\text{ABP}}$. For example, for $\text{Pe}_r = 160$ the minimum at $ p = 0.4$ amounts to $T / T^{\text{ABP}} = 0.18$. Interestingly, this minimum occurs at a dipole strength comparable to the one estimated for \textit{E.~coli} bacteria  \cite{Drescher11}.

 In region II, $T$ grows to $10 T^{\text{ABP}}$ or well beyond. The orientation of the pusher has time to equilibrate about $\theta_s = \pi/2$ and then attempts to reach $\theta^{*}$ by rotational noise. Indeed, one can rewrite the effective rotational drift in Eq.\ (\ref{Ref-ADJFPE1D}) by introducing an effective angular potential $\Omega = - \partial V_{\text{eff}} / \partial \theta$ with $V_{\text{eff}} = V + V_r = 3 p \cos^2\theta/16 - \ln(\sin\theta)/(2\text{Pe}_r)$, where the second term comes from the 3D rotational diffusion. However, the pusher escaping from the wall at $\theta^{\ast}$ cannot be viewed as a typical Kramers problem \cite{Honerkamp90} since the orientation vector $\mathbf{e}$ does not pass a smooth potential barrier of height $\Delta V_{\mathrm{eff}}$ when reaching the escape angle $\theta^{\ast}$. Instead, the swimmer orientation moves up the potential $V_{\text{eff}}$ by an amount $\Delta V_{\mathrm{eff}} = V_{\mathrm{eff}}(\theta^{*})- V_{\mathrm{eff}}(\theta_s)$ and when the pusher leaves the wall at $\theta^{*}$, it also leaves the range of $V_{\text{eff}}$ [see Fig.\ \ref{Fig:Fig3}(b)]. However, we can derive an approximate formula for large $\mathrm{Pe}_r \Delta V_{\mathrm{eff}}$ with the Arrhenius factor reminiscent of Kramers' mean escape time  \cite{supp,Joern11},
\begin{align}
T^{\text{pusher}} \approx  \frac{\sqrt{\pi}}{|V_{\text{eff}}'(\theta^{*})| \sqrt{ \text{Pe}_r V_{\text{eff}}''(\theta_s)}} e^{2\text{Pe}_r \Delta V_{\mathrm{eff}}} \, .
\label{Eq:TPush}
\end{align}
Interestingly, in case of the puller, the rotational-noise contribution $V_r$ shifts the most stable orientation to $\theta_s = \pi - \arcsin[2/ \sqrt{-3p\,\text{Pe}_r}] < \pi$ [see Fig.\ \ref{Fig:Fig3}(c)] \cite{supp}. Here, we can approximate $T$  by Kramers' formula \cite{supp,footnote1}
\begin{align}
T^{\text{puller}} \approx  \frac{\pi}{ \sqrt{ |V_{\text{eff}}''(\theta^{*})| V_{\text{eff}}''(\theta_s)}} e^{2\text{Pe}_r \Delta V_{\mathrm{eff}}} \, .
\label{Eq:TPull}
\end{align}
The inset of Fig.~\ref{Fig:Fig3}(a) demonstrates that $T$ calculated from Eqs.~(\ref{Eq:TPush}) and (\ref{Eq:TPull}) at $|p| \mathrm{Pe}_r \gg 1$ agrees very well with the one obtained by numerically solving  Eqs.~(\ref{Ref-ADJFPE1D}) and (\ref{Relation-f-and-g}).

While so far we considered generic microswimmer models, we now turn to the spherical  \textit{squirmer} \cite{squirmer}, which serves as a model for ciliated microorganisms such as \textit{Paramecium} \cite{squirmer,Ishikawa06b}  and \textit{Volvox} \cite{Drescher09} but also for active emulsion droplets\ \cite{Thutupalli11}. The squirmer propels itself by an axisymmetric surface velocity field $  \mathbf{v}_s =   \frac 3 2 \left( 1 + \beta \mathbf{e}\cdot \hat{\mathbf{r}}_{s} \right)   \left[ (\mathbf{e}\cdot \hat{\mathbf{r}}_{s} ) \hat{\mathbf{r}}_{s} - \mathbf{e}  \right]$, where  $\hat{\mathbf{r}}_{s}$ is the unit vector pointing from the center of the squirmer to its surface. The neutral squirmer ($\beta=0$) creates the bulk flow field of a source dipole with $q = 1/2 $, while $\beta \ne 0$ adds an additional force-dipole field with $p = - 3\beta /4 $ \cite{Ishikawa06}. Recent studies with squirmer-wall interactions already exist but without any noise \cite{Llopis10,Spagnolie12,Ishimoto13,Li14}. Using lubrication theory, the authors of Ref.\ \cite{Ishikawa06} have calculated the  dimensionless friction torque acting on the squirmer in front of a wall due to hydrodynamic interactions \cite{Ishikawa06},
\begin{align}
 M =  (6\pi/5) (1-\beta\cos\theta) \sin\theta(\ln\epsilon^{-1} - c),
\label{Eq-LubricationSq} 
\end{align}
where $\epsilon =h-1\! \ll\! 1$ is a small distance and $c\!=\!\text{const}$. This gives the wall-induced angular velocity $\Omega_{\text{HI}}=-  M /\gamma_r$, where $\gamma_r$ is the rotational friction coefficient near the surface \cite{Cichocki98,fn2}. Note that the neutral squirmer ($\beta=0$) behaves like the generic source dipole even  close to the wall since $\Omega_{\text{HI}} \sim -\sin\theta$. This might explain why far-field hydrodynamic interactions describe the near-wall swimming of neutral squirmers as shown in \cite{Spagnolie12}. The $\beta$-dependent part in Eq.~(\ref{Eq-LubricationSq}) adds to $\Omega_{\text{HI}}$ the force-dipole term $\sim - p \sin\theta \cos\theta$. Acting alone, it rotates the squirmer pusher ($\beta < 0$) towards the wall and therefore it behaves like the generic puller with increased detention time and vice versa. These results are in accordance with recent simulations at finite Reynolds numbers \cite{Li14}.

\begin{figure}[t]
\includegraphics[width=0.86\columnwidth]{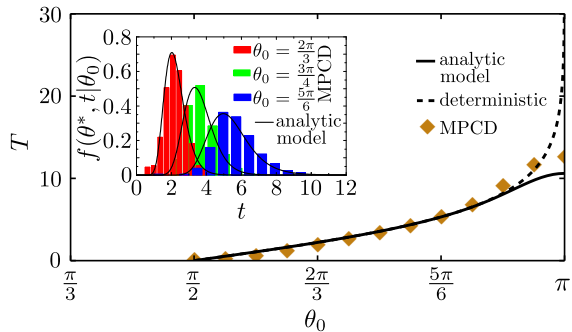}
\caption{
Mean detention time $T$ of a neutral squirmer plotted versus the initial angle $\theta_{0}$ for $\text{Pe}_{r}=110$ and $\epsilon=0.01$ (approximate mean distance from the wall measured from MPCD simulations) and compared to the analytic 1D model [Eqs.~(\ref{Ref-ADJFPE1D}) and (\ref{Relation-f-and-g})], and the deterministic model ($\text{Pe}_r \rightarrow\infty$). Inset: Distribution of detention times from MPCD simulations and compared to  analytic model.}
\label{Fig:Fig4ab}
\end{figure}

To demonstrate that our 1D model is applicable, we perform full 3D mesoscale hydrodynamic simulations using multi-particle collision dynamics (MPCD)\cite{Malevanets99,Kapral08,Gompper08}. It solves the Navier-Stokes equations for the fluid around the squirmer and the wall and naturally includes thermal fluctuations\ \cite{Downton09,Goetze10,Zoettl12,Zoettl14}. First, we numerically determine $c\approx 0.9$ \cite{supp} and then explicitly simulate many swimming trajectories of swimmer-wall collision events for a neutral squirmer at different incoming angles. Figure~\ref{Fig:Fig4ab} shows results for the mean detention time $T$ plotted versus  the initial angle $\theta_0$, which agree well   with our analytic model. The mean detention time of the deterministic swimmer, $T^{\text{det}} \propto \ln\tan(\theta_0/2)$ \cite{supp}, deviates from the full model only close to the unstable equilibrium orientation at $\theta = \pi$. Here $T^{\text{det}} \rightarrow \infty$, whereas noise renders $T$ finite and helps the swimmer to escape. The inset of Fig.~\ref{Fig:Fig4ab} shows a convincing agreement of the DTDs determined from the analytic model and MPCD simulations.

To assess fluctuations of the position $h(t)$ above the surface, which influence $\Omega_{\text{HI}}$ \cite{Berke08,Li08}, we may define an escape event by reaching a certain escape height $h^{*}>1$. For the state variable $\mathbf{y}(t)=(h,\theta)$ one defines the probability $g(\mathbf{y}^*,t | \mathbf{y}_0)$, for finding the swimmer below  $h^{*}$ at time $t =t^*-t_0$ while the initial state $\mathbf{y}_0$  at $t_{0}$ starts at $h_{0}\in[1,h^{*})$ and $\theta_{0}\in[0,\pi]$\ \cite{Schaar13}. The probability obeys the adjoint Fokker-Planck equation 
\begin{equation}
\begin{split}
& \partial_{t} g(\mathbf{y}^{*},t | \mathbf{y}_0) = [(\mathbf{v}_{\text{A}}+\mathbf{v}_{\text{HI}})\cdot \mathbf{e}_{z}\partial_{h_{0}}
 +D_{t}\partial^{2}_{h_{0}} \\
&-  (\Omega_{\text{HI}} +D_r\cot\theta_0)\partial_{\theta_{0}}+D_{r}\partial^{2}_{\theta_{0}}]g(\mathbf{y}^{*},t|\mathbf{y}_0)
\label{Ref-ADJFPE2D},
\end{split}
\end{equation}
with the initial condition $g(\mathbf{y}^{*},t_{0}|\mathbf{y}_0)= \delta(\mathbf{y}^{*}-\mathbf{y}_0)$,  and reflecting [at $\mathbf{y}_0 = (1,\pi)$] and absorbing [at $\mathbf{y}_0 = (h^*,\theta^*)$] boundary conditions for $g(\mathbf{y}^*,t | \mathbf{y}_0)$. Then,  $f(\mathbf{y}^*,t | \mathbf{y}_0) = -\partial_t g(\mathbf{y}^*,t | \mathbf{y}_0)$ is the DTD for detention time $t$. In \cite{supp} we show that for sufficiently large Pe and $h^{*}$ the detention times in the 2D model are larger compared to the 1D model. Small $h^{*}$ can also be reached by translational Brownian motion which reduces the detention times.

To conclude, based on the method of first-passage times, we developed a formalism to determine the distribution of detention times for microswimmers near a plane no-slip surface taking into account hydrodynamic interactions and rotational noise. For generic microswimmers such as source dipoles, pushers, and pullers we demonstrated that the mean detention time can vary over several orders of magnitude relative to the ABP depending on persistence number $\mathrm{Pe}_r$ and swimmer strengths $q$, $p$. This allows us to quantify the relative importance of hydrodynamic interactions and rotational noise. Our model also provides a route to quantify  wall accumulation of microswimmer suspensions confined between two plates, as determined experimentally for different  microorganisms  \cite{Rothschild63,Berke08,Li09,Molaei14}. Our method can be extended to include further drift terms, for example, due to non-spherical swimmer shape, which further modifies the reorientation dynamics at the wall \cite{Li09,Spagnolie12}. Therefore, it offers a systematic approach for studying how artificial as well as biological microswimmers behave at surfaces.

We thank Giovanni Volpe and Katrin Wolff for helpful discussions and the DFG for support within the research training group GRK 1558, within the priority program SPP 1726 "Microswimmers" (STA 352/11), and by grant STA 352/10-1.

\newpage
\textcolor{white}{.}
\newpage
\textbf{SUPPLEMENTAL MATERIAL} \\[1mm]
\noindent\textbf{1.~Adjoint Smoluchowski equation} \\

The dynamics of the conditional probability density $P(\mathbf{e},t^{\ast}|\mathbf{e}_0,t_0)=P(\theta,\phi,t^{\ast}|\theta_0,\phi_0,t_0)$ for finding a swimmer in the orientation interval
 $[\mathbf{e},\mathbf{e}+\mathrm{d}\mathbf{e}]$ at time $t^{*}$,
if it was at orientation $\mathbf{e}_0$ at time $t_0$, is governed by the Smoluchowski equation
\begin{align}
\partial_{t^{\ast}} P(\mathbf{e},t^{\ast}|\mathbf{e}_0,t_0)  = \mathcal{L}(\mathbf{e})   P(\mathbf{e},t^{\ast}|\mathbf{e}_0,t_0) \, ,
\label{Eq:LFor1}
\end{align}
where 
\begin{align}
 \mathcal{L}(\mathbf{e})  = -\mathcal{R}(\mathbf{e}) \cdot \boldsymbol{\Omega}_{\text{HI}}(\mathbf{e}) + D_r\mathcal{R}^2(\mathbf{e})
\label{Eq:LFor2}
\end{align}
is the Smoluchowski operator
and $\mathcal{R}(\mathbf{e})=\mathbf{e} \times \boldsymbol{\nabla}_{\mathbf{e}}$  the rotation operator (see main text).
Eq.~(\ref{Eq:LFor1}) describes the \textit{forward} evolution of the probability density $P(\mathbf{e},t^{\ast}|\mathbf{e}_0,t_0)$
as a function of $\mathbf{e}$ and $t^{\ast}$ for the initial condition $P(\mathbf{e},t_0|\mathbf{e}_0,t_0)=\delta(\mathbf{e}-\mathbf{e}_0)$ at $t^{\ast}=t_0$.
One can also formulate the \textit{backward} evolution of $P(\mathbf{e},t^{\ast}|\mathbf{e}_0,t_0)$,
now as a function of $\mathbf{e}_0$ and $t_0$, which is described by the adjoint Smoluchowski equation (see, e.g.,~\cite{Honerkamp90} for more details) 
\begin{align}
\partial_{t_0} P(\mathbf{e},t^{\ast}|\mathbf{e}_0,t_0)  = -\mathcal{L}^{+}(\mathbf{e}_0)   P(\mathbf{e},t^{\ast}|\mathbf{e}_0,t_0) \, ,
\label{Eq:LBack1}
\end{align}
where 
\begin{align}
 \mathcal{L}^{+}(\mathbf{e}_0)  = \boldsymbol{\Omega}_{\text{HI}}(\mathbf{e}_0) \cdot \mathcal{R}(\mathbf{e}_0) + D_r\mathcal{R}^2(\mathbf{e}_0)
\label{Eq:LBack2}
\end{align}
is the adjoint operator of  $\mathcal{L}(\mathbf{e}_0)$.
In our problem $\boldsymbol{\Omega}_{\text{HI}}(\mathbf{e}_0)= \Omega_{\text{HI}}(\theta_0)\mathbf{e}_{\phi_0}$ (see main text)
and so $\boldsymbol{\Omega}_{\text{HI}}(\mathbf{e}_0)  \cdot \mathcal{R}(\mathbf{e}_0) = \Omega_{\text{HI}}(\theta_0)\partial_{\theta_0}$
is independent of $\phi_0$.
Taking now the integral $\int_{0}^{2\pi} \mathrm{d}\phi \int_{0}^{2\pi} \mathrm{d}\phi_0\dots$ of Eq.~(\ref{Eq:LBack1}) yields
\begin{align}
\partial_{t_0} p(\theta,t^{\ast}|\theta_0,t_0)  = -\mathcal{L}^{+}(\theta_0)   p(\theta,t^{\ast}|\theta_0,t_0) \, 
\label{Eq:LBack3}
\end{align}
with $p(\theta, t^{\ast} | \theta_0,t_0)= \int_{0}^{2\pi} \mathrm{d}\phi_0 \int_{0}^{2\pi} \mathrm{d}\phi P(\theta, \phi, t^{\ast} | \theta_0,\phi_0,t_0)$
and $\mathcal{L}^{+}(\theta_0) = \Omega_{\text{HI}}(\theta_0)\partial_{\theta_0} + D_r(\partial^2_{\theta_0} + \cot \partial_{\theta_0})$
(see main text).
Finally, taking the integral $\int_{\theta^{\ast}}^{\pi}\dots\sin\theta\mathrm{d}\theta$ of Eq.~(\ref{Eq:LBack3})
and introducing $t=t^{\ast}-t_0$
 results in
Eq.~(2) of the main text.

\vspace{5mm}\noindent\textbf{2.~Estimate of Mean Detention Time} \\

In the following, we derive the estimates for the mean detention times of pushers and pullers [see Eqs. (6) and (7) of
main text], when hydrodynamic interactions with the surface are sufficiently large. We will consider the escape from the surface
as an escape process from the minimum of an effective potential and formulate equations reminiscent of Kramers'
mean escape rate\ \cite{Honerkamp90}.
The dimensionless adjoint Smoluchowski equation for  $g( \theta^{\ast}, t | \theta_{0} )$  [see Eq.~(2) of the main text] can be 
rewritten
as
\begin{align}
\frac{\partial g( \theta^{\ast}, t | \theta_{0} )}{\partial t}  = \left[
  \frac{1}{2 \text{Pe}_r}\frac{\partial }{\partial  \theta_0} -\frac{\partial  \Ueff(\theta_0)}{\partial \theta_0}
\right]
\frac{\partial   g( \theta^{\ast}, t | \theta_{0})}{\partial  \theta_0}  \, ,
%
%
\end{align}
where the effective potential $\Ueff(\theta)$ for the force-dipole swimmers 
reads
$
\Ueff(\theta) = V + V_r = 3p\cos^2\theta / 16  - \ln(\sin\theta)/(2\Per)
$.
The stable orientations $\theta_s$ 
of
the pusher and the puller are determined by the potential 
minimum of $\Ueff(\theta_s)$,
where 
$\left. \Ueff'(\theta)\right|_{\theta=\theta_s}=0$ and $\left. \Ueff''(\theta)\right|_{\theta=\theta_s}>0$.
The stable orientation 
of
the pusher reads $\theta_s=\pi/2$  
and the puller orients at
$\theta_s=\pi -  \arcsin[2/ \sqrt{-3p\text{Pe}_r}]$. 
As explained in the main text, the
escape angle of the pusher is $\theta^{*} = \arccos[(-4+\sqrt{16+27p^{2}})/(9p)]$ and for the puller
$\theta^{*} = \pi/2$. 

In the following we assume that the angular dynamics $\theta(t)$ can be separated into two processes: 
First, after reaching the wall at the incoming angle $\theta_0$, the swimmer is oriented by hydrodynamic interactions 
towards the stable orientation $\theta_s$ at the wall.
There, the swimmer orientation equilibrates fast in the minimum and
a (quasi-) stationary distribution $p(\theta)$  peaked around $\theta_s$ is established.
All in all, this takes the typical time $\tau(\theta_0 \rightarrow \theta_s)$.
Second, starting from $\theta_s$ 
the swimmer tries to escape from the wall. For reaching 
the escape angle $\theta^{*}$, it has to move up the potential difference $\Delta \Ueff = \Ueff(\theta^{*}) - \Ueff(\theta_s)$, which
takes the time $\tau(\theta_s \rightarrow \theta^{*})$.
Then, the mean first-passage time from the wall can be approximated by $T \approx \tau(\theta_0 \rightarrow \theta_s) + \tau(\theta_s \rightarrow \theta^{*})$.
We now assume that $\tau(\theta_s \rightarrow \theta^{*}) \gg \tau(\theta_0 \rightarrow \theta_s)$ and hence $T \approx \tau(\theta_s \rightarrow \theta^{*})$.
So, we can approximate the escape process from the surface as a quasi-stationary dissociation process
to move up the potential difference
$\Delta \Ueff$ by starting at the potential minimum $\theta_s$.

Now, the theory of mean first-passage times provides an exact expression for the mean escape time from the wall,
calculated as the mean time needed for reaching
a potential value $\Ueff(\theta^{*})$ at $\theta^{*}$  when starting from $\Ueff(\theta_s)$ at $\theta_s$
\cite{Honerkamp90}:
\begin{align}
T  
  =  2\text{Pe}_r \int_{\theta^{*}}^{\theta_s}d\theta_1 e^{2\text{Pe}_r \Ueff(\theta_1)}  \left( \int_{\theta_1}^{\pi}d\theta_2 e^{-2\text{Pe}_r \Ueff(\theta_2)} \right).
\label{Eq:T1}
\end{align}
Eq.~(\ref{Eq:T1}) cannot be solved exactly.
However, for $\text{Pe}_r \Delta \Ueff \gg 1$
the first integral on the right-hand side of Eq.\ (\ref{Eq:T1}) is dominated by the maximum potential
value at $\theta^{*}$ so that $T$ can be approximated 
by
\cite{Honerkamp90}
\begin{align}
T 
 \approx  2\text{Pe}_r \int_{\theta^{*}}^{\theta_s}d\theta_1 e^{2\text{Pe}_r \Ueff(\theta_1)}  \int_{\theta^{*}}^{\pi}d\theta_2 e^{-2\text{Pe}_r \Ueff(\theta_2)} \, .
\label{Eq:T2}
\end{align}
Now, the two independent integrations are mainly governed by the regions around the respective
maximum values of the integrands at  $\theta^{*}$  and $\theta_s$.


\vspace{5mm}\noindent\textsl{2.1~Pusher} \\

For 
smooth potential barriers $T$  can be calculated by using Kramers' escape-time formula \cite{Honerkamp90}.
However, it is not applicable to the escape of a pusher from the wall. The reason is that the pusher does not pass 
a potential maximum of $\Ueff$ but only has to reach
 $\theta^{*}$ at the slope of the potential,
where it ultimately escapes.  
An approximate solution of Eq.~(\ref{Eq:T2}) can be found by expanding the potential around $\theta^{*}$ for the $\theta_1$ integration
and around $\theta_s$ for the $\theta_2$ integration in Eq.~(\ref{Eq:T2}).
So, integration over $\theta_2$ yields
\begin{align}
\begin{split}
&  \int_{\theta^{*}}^{\pi}d\theta_2 e^{-2\text{Pe}_r \Ueff(\theta_2)} \\
 \approx &  \int_{-\infty}^{\infty}d\theta_2 e^{-2\Per [\Ueff(\theta_s) + \frac{1}{2}\Ueff''(\theta_s)(\theta_2-\theta_s)^2]} \\
 = &  e^{-2\text{Pe}_r \Ueff(\theta_s)} \sqrt{\frac{\pi}{ \Per \Ueff''(\theta_s)}}         = \sqrt{\frac{2\pi}{ \frac{3}{4}p\text{Pe}_r + 1 }} \, ,
\end{split}
\label{Eq:T3a}
\end{align}
where we used $ \Ueff''(\theta_s) =  3p/8 + (2\text{Pe}_r)^{-1}$ and $\Ueff(\theta_s)=0$,
and integration over $\theta_1$ is evaluated to
\begin{align}
\begin{split}
  \int_{\theta^{*}}^{\theta_s}d\theta_1 e^{2\text{Pe}_r \Ueff(\theta_1)} &  \approx  \int_{\theta^{*}}^{\infty}d\theta_1  e^{2\Per [\Ueff(\theta^{*}) + \Ueff'(\theta^{*})(\theta_1-\theta^{*})]} \\
  = \frac{e^{2\Per\Ueff(\theta^{*})}}{2\Per |\Ueff'(\theta^{*})|}       & = \frac{1}{2\text{Pe}_r \Omega (\theta^{*})} e^{2\text{Pe}_r \Delta \Ueff}
\, ,
\end{split}
\label{Eq:T3b}
\end{align}
where we used $\Ueff'(\theta^{*}) = - \Omega (\theta^{*})$ (see main text) and $ \Delta \Ueff = \Ueff(\theta^{*})$.
Hence, 
with Eqs.~(\ref{Eq:T2}), (\ref{Eq:T3a}) and  (\ref{Eq:T3b}) we obtain Eq.~(6) of the main text.


\vspace{5mm}\noindent\textsl{2.2~Puller} \\

The escape of the puller ($p<0$) at the location $\theta^{*} = \pi/2$ of the potential maximum $U(\theta^{*})=0$ 
starting from the potential minimum $U(\theta_s)<0$ 
can be calculated by using a harmonic expansion 
around both the potential minimum and 
the potential maximum.
So, the integrals in Eq.~(\ref{Eq:T2}) can be approximated 
by
\begin{align}
\begin{split}
 &\int_{\theta^{*}}^{\pi}d\theta_2 e^{-2\text{Pe}_r \Ueff(\theta_2)} \\
 \approx &  \int_{-\infty}^{\infty}d\theta_2 e^{-2\Per [\Ueff(\theta_s) + \frac{1}{2}\Ueff''(\theta_s)(\theta_2-\theta_s)^2]} \\
  = & \frac{\sqrt{\pi} e^{-2\text{Pe}_r \Ueff(\theta_s)}    }{\sqrt{\Per \Ueff''(\theta_s)}}     = \sqrt{\frac{\pi}{ \frac{3}{4}|p|\text{Pe}_r - 1 }}e^{2\text{Pe}_r \Delta \Ueff} \, ,
\end{split}
\label{Eq:T4a}
\end{align}
where we used $\Delta \Ueff =  |\Ueff(\theta_s)|$ and $\Ueff''(\theta_s) = -3p /4 - 1/\text{Pe}_r $,
and by
\begin{align}
\begin{split}
 &  \int_{\theta^{*}}^{\theta_s}d\theta_1 e^{2\text{Pe}_r \Ueff(\theta_1)}  \\
 \approx &  \int_{\theta^{*}}^{\infty}d\theta_1  e^{2\Per [\Ueff(\theta^{*}) + \frac 1 2 \Ueff''(\theta^{*})(\theta_1-\theta^{*})^2]} \\
  = & \frac 1 2 \sqrt{\frac{\pi}{\Per |\Ueff''(\theta^{*})|}} e^{2\Per \Ueff (\theta^{*})}  = \frac 1 2 \sqrt{\frac{2 \pi}{ \frac{3}{4}|p|\text{Pe}_r - 1 }}
\end{split}
\label{Eq:T4b}
\end{align}
where we used $\Ueff''(\theta^{*}) =  3p/8 + 1/(2 \text{Pe}_r)$ and $\Ueff (\theta^{*}) = 0$.
Hence, 
with Eqs.~(\ref{Eq:T2}), (\ref{Eq:T4a}) and  (\ref{Eq:T4b}) we obtain Eq.~(7) of the main text.

We note that in the limit $|p|\Per \rightarrow \infty$ the potential around the minimum at $\theta_s \rightarrow \pi$ is 
highly asymmetric and the quadratic expansion is 
not a good approximation.
For this case we assume that the stable position is $\theta_s \approx \pi$.   
Then, we approximate the $\theta_2$-integration in Eq.~(\ref{Eq:T2}) for the puller
by
\begin{equation}
\begin{split}
 & \int_{-\infty}^{\theta_s}d \theta_2 e^{-2\Per\Ueff(\theta_2)}  = \int_{-\infty}^{\theta_s}d\theta_2 e^{-2\Per V(\theta_2)}e^{-2\Per V_r(\theta_2)} \\
    \approx &  e^{-2\Per V(\theta_s)} \int_{-\infty}^{\theta_s}d\theta_2 \sin \theta_2  e^{-\Per V^{''}(\theta_s)(\theta_2-\theta_s)^2} \\
    \approx &   - e^{-2\Per V(\theta_s)} \int_{-\infty}^{\theta_s}d\theta_2  (\theta_2- \theta_s)  e^{-\Per V^{''}(\theta_s)(\theta_2-\theta_s)^2} \\
    = &   \frac {e^{-2\Per V(\theta_s)}}{2 \Per V^{''}(\theta_s) } =   \frac{4e^{2\Per \Delta V}}{3|p|\Per}
\label{TPull1}
\end{split}
\end{equation}
and together with Eq.~(\ref{Eq:T4b})
we obtain
\begin{align}
T(\theta_s \rightarrow \theta^{*}) \approx   \frac{1}{2}  \frac{ \sqrt{\pi} e^{2\Per \Delta V}}{\sqrt{\Per |V^{''}(\theta^{*})|}\,\, V^{''}(\theta_s) } 
\label{T5}
\end{align}
with  $V^{''}(\theta^{*}) = \frac{3p}{8}$ and $V^{''}(\theta_s) = -\frac{3p}{8}$.

\vspace{5mm}\noindent\textsl{2.3~Deterministic Detention Times} \\

For the deterministic source-dipole swimmer $\Omega_{\text{HI}} = d\theta / dt = - \frac{3q}{8}\sin\theta$ in dimensionless units.
After integration we obtain for an incoming angle $\theta_0$ and an escape angle $\theta^{*}=\pi/2$  the deterministic wall detention time
\begin{align}
T^{\text{det}} = \frac{8}{3q} \ln (\tan \frac{\theta_0}{2}).
\label{Eq:DetSD}
\end{align}
For a neutral squirmer close to a wall
$\Omega_{\text{HI}} = -M/ \gamma_r = - \frac{6\pi}{5}\frac{\ln \epsilon^{-1} - c}{\gamma_r(\epsilon)}\sin\theta$
where we use $M$ from Eq.~(8) of the main text and $\gamma_r \approx 8\pi [(2/5) \ln \epsilon^{-1} + 0.37]$ is the dimensionless rotational friction constant
of a sphere near a wall \cite{Cichocki98}.
Hence
\begin{align}
T^{\text{det}} = \frac{8}{3}\frac{(\ln \epsilon^{-1} + 0.925)}{(\ln \epsilon^{-1} - c)}  \ln (\tan \frac{\theta_0}{2}).
\label{Eq:DetSD2}
\end{align}


The noiseless pusher and puller have stable orientations at the wall and hence $T^{\text{det}} \rightarrow \infty$.


\vspace{5mm}\noindent\textbf{3.~Additional Information on the Motion of a Swimmer Near a Wall} \\

\noindent\textsl{3.1~Trajectory in Front of a Wall} \\


\begin{figure}
\includegraphics[width=0.8\columnwidth]{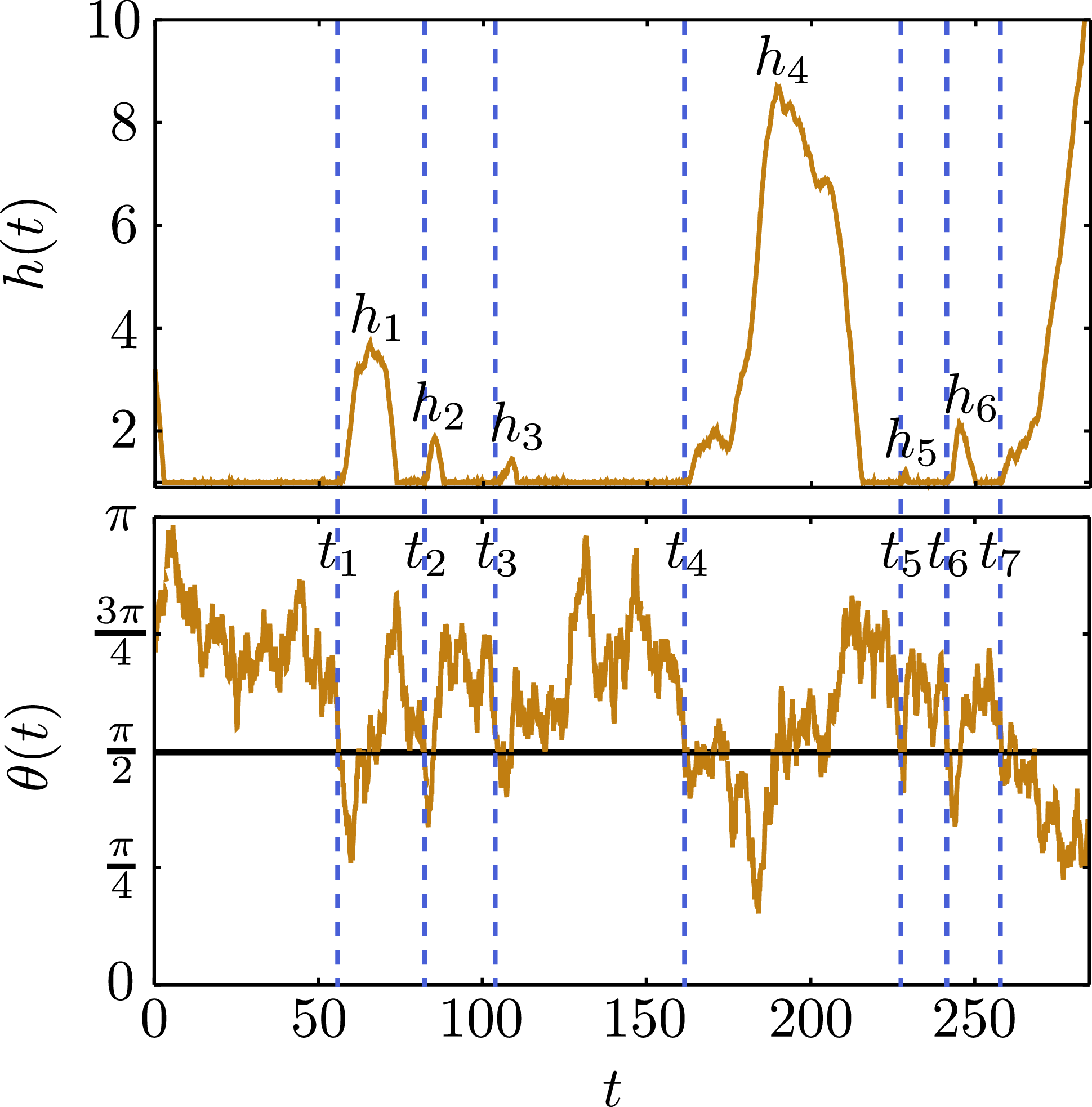}
\caption{Typical trajectory of a swimmer close to a wall. The time evolution of the swimmer-wall distance $h(t)$ and the orientation $\theta(t)$ is shown here for an active Brownian particle with $\text{Pe}_r=25$, $\text{Pe}=1000$.
At times $t_n$ the swimmer reaches the escape angle $\theta^{*}$.
Between two encounters the swimmer approaches a maximum distance $h_n$ above the wall.
}
\label{Fig:traj}
\end{figure}

Swimmer trajectories can be obtained by numerically solving Eqs.~(1) of the main text.
In Fig.~\ref{Fig:traj} we show an example for an ABP (with $\text{Pe}_r=25$, $\text{Pe}=1000$).
The swimmer starts at an angle $\theta \approx 3\pi/4$ at a distance $h\approx 3$ from the wall, approaches the wall, reorients, and its position $h(t)$ fluctuates.
When the swimmer reaches an escape angle $\theta^{*} = \pi/2$ at times $t_n$, $n$=$1,\dots,7$, it  swims away from the wall.
Due to the persistent random walk of the swimmer it will come back to the wall infinitely many times.
We only show the trajectory here until $t\approx 280$.
The maximum heights above the surface $h_n$, $n=1,\dots,6$, between two encounters with a wall strongly fluctuate.

Hence, defining the escape of a swimmer from a surface by introducing a specific escape height  $h^{*}$, strongly depends on the value of  $h^{*}$.
A much clearer and unambiguous definition of the escape process is our approach in the main text, where we introduce an escape angle within a 1D model.
The resulting mean detention time is an appropriate means to compare the
swimmer-wall encounters of different swimmer types to each other.

\vspace{5mm}\noindent\textsl{3.2~Orientational distribution at escape height $h^{*}>1$} \\

\begin{figure}
\includegraphics[width=0.9\columnwidth]{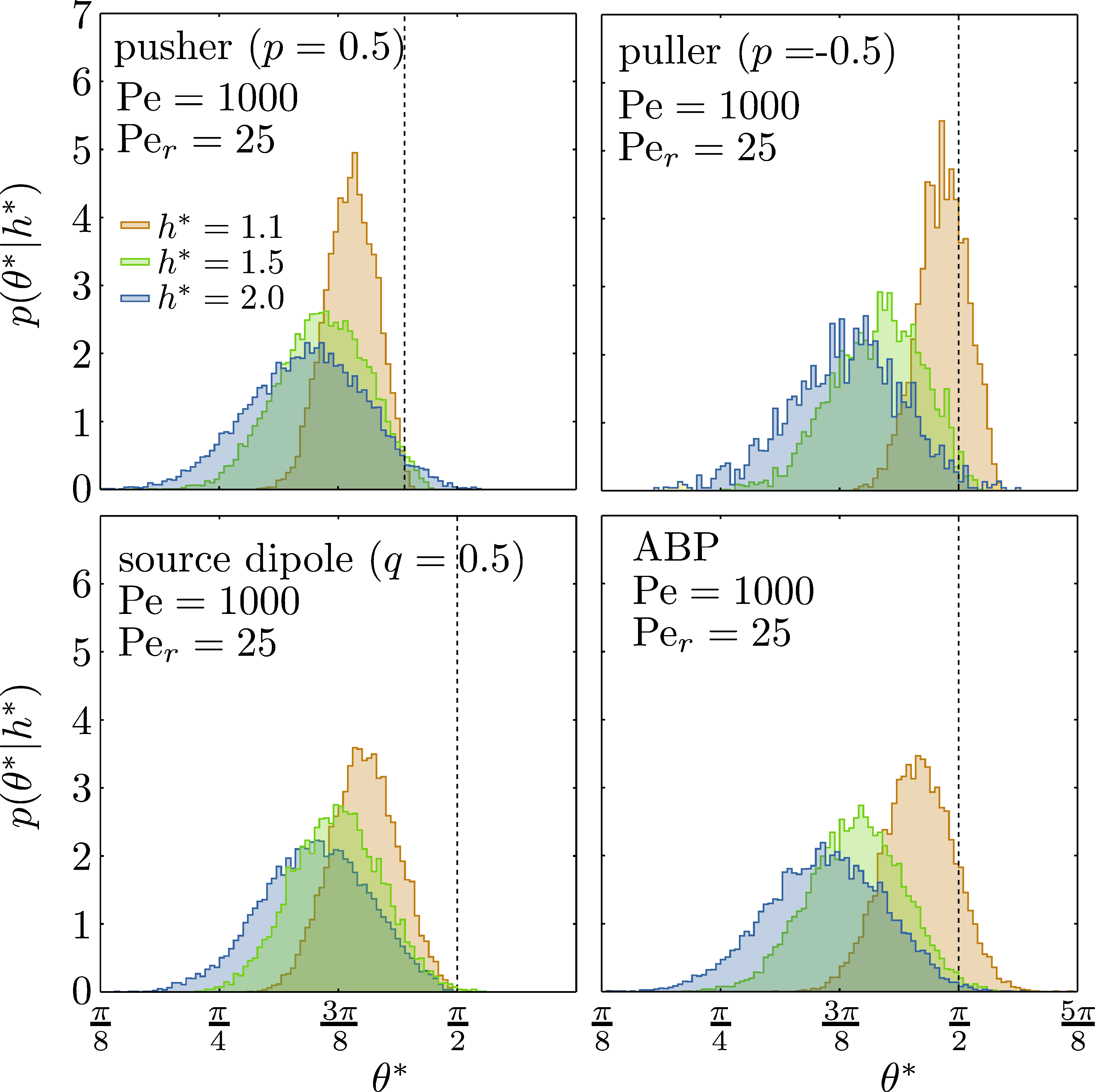}
\caption{Distribution of angles $\theta^{*}$ after collision with a surface for different swimmer types measured at several distances $h^{*}$ above the wall.
The dashed lines show the escape angles from the 1D model.}
\label{Fig:thetaout}
\end{figure}

One may define the escape of the swimmer from the wall by reaching 
 a certain height $h^{*}>1$ above the wall after leaving the wall, as suggested in our 2D model [Eq.(9) in the main text].
Then the angles $\theta^{*}$ at $h^{*}$ are typically smaller than the escape angles defined in the 1D model (see main text)
and are distributed over a range of $\theta^{*}$ values.
Some distributions $p(\theta^{*}|h^{*})$ are shown in Fig.~\ref{Fig:thetaout} for different swimmer types for $\text{Pe}_r=25$, $\text{Pe}=1000$.
The widths of the distributions typically increase and the mean values $\langle \theta^{*} \rangle$ decrease with increasing $h^{*}$, since the swimmers have more time to explore a larger range of orientations.
Since swimmers can also reach the escape height $h^{*}$ by translational Brownian motion, the corresponding
orientation angle $\theta^{*}$ can also be larger than the escape angle needed in the 1D model  (dashed lines).
Nevertheless, for sufficiently large $h^{*}$ escape through translational diffusion becomes negligible, as also shown in Sec.~3.3 and 3.4.

\vspace{5mm}\noindent\textsl{3.3~Translational vs.~Orientational Escape Times} \\

When a swimmer approaches a wall, it is subjected to both translational and rotational noise, quantified by the P\'eclet number $\text{Pe}$ and the persistence number  $\text{Pe}_r$  (see definitions in the main text).
For a spherical swimmer with pure thermal noise, translational and rotation diffusion constants are coupled \cite{Enculescu11} and $\text{Pe}=3\text{Pe}_r$.
For real microswimmers intrinsic noise can enhance rotational diffusion and thereby lower its persistence, such that
\begin{equation}
\text{Pe} \ge 3\text{Pe}_r.
\label{Eq:PePer}
\end{equation}
In the main text we presented the mean detention times $T$ at the wall by assuming the swimmer stays at the wall ($h\approx R$)
until it reaches the escape angle $\theta^{*}$.
In the following we  show that our model swimmers indeed stay close to the wall during time $T$ and despite their translational diffusion.
We estimate a translational escape time $T_t(\Delta h)$ for reaching a height $\Delta h$ (in units of the swimmer radius $R$) above the surface.
By setting $T_t = T$, we calculate the height $\Delta h$, the swimmer can reach in time $T$. We find $\Delta h<R$, so the swimmer stays close to the surface.

Close to the wall the swimmer velocity along the surface normal  in units of the swimmer velocity $U$ is $v_W=\cos\theta + v_{\text{HI}}(\theta,h)$ due to active motion  and hydrodynamic swimmer-wall interactions.
For a constant drift velocity $v_W<0$ and a given P\'eclet number,
one can readily solve Eq.~(9) from the 
main text following Ref.~\cite{Redner}.
In particular, the mean time
 $T_t$ to move away a distance $\Delta h$ from the wall after starting at the wall reads  in dimensionless units 
\begin{equation}
T_t(\Delta h, v_W,\text{Pe}) =  \frac{\exp(2\text{Pe}\Delta h |v_W|)-1}{2 \text{Pe}v_W^2} -  \frac{\Delta h}{|v_W|}.
\label{Eq:Tt}
\end{equation}
For the limiting cases $\text{Pe}\rightarrow\infty$ and $\text{Pe}\rightarrow 0$ one obtains the expressions
\begin{equation}
\begin{split}
T_t(\text{Pe}\rightarrow\infty) = & \frac{ \exp(2\text{Pe}\Delta h|v_W|)}{2\text{Pe}v_W^2},\\
T_t(\text{Pe}\rightarrow 0) = & \text{Pe} \Delta h^2.  
\label{Eq:Tt2}
\end{split}
\end{equation}
We will use Eqs.~(\ref{Eq:Tt}) and (\ref{Eq:Tt2}) in the following 
to estimate lower limits for the time $T_t$ and hence the maximum distance  $\Delta h$,
which a swimmer reaches during reorienting at the wall.


\vspace{4mm}\noindent\textsl{3.3.1~Active Brownian Particle (ABP)} \\[-2mm]

By rotational diffusion an ABP without hydrodynamic swimmer-wall interactions  needs  $T^{\text{ABP}} < 2 \text{Pe}_r$ to reach the escape orientation for any incoming angle $\theta_0$, as shown in Eq.~(5) of the main text.
On the other hand, Eq.~(\ref{Eq:Tt}) in the limit $v_w \rightarrow 0$, which means parallel orientation, predicts a lower limit for $T_t$ to reach a height $\Delta h$ above the surface: $T_t^{\text{ABP}} > \text{Pe} \Delta h^2$.
By setting $T\sim T_t$, we estimate an upper limit for the height ($\Delta h^{\text{ABP}}<\sqrt{2/3}$) the ABP reaches by translational diffusion during reorientation at the wall.
Hence, the escape process for the ABP is determined by rotational diffusion at all P\'eclet numbers.
Since for the ABP the reorientation rate does not depend on $h$, the detention times are not modified by translational diffusion.


\vspace{4mm}\noindent\textsl{3.3.2~Pusher} \\[-2mm]

Figure~3 in the main text shows the detention times $T$ for force dipole swimmers depending on the dipole strength $p$ and the persistence number $\text{Pe}_r$.
For the pusher ($p>0$) in region I we have $T<T^{\text{ABP}}$ (see main text).
To obtain estimates for the translational escape times $T_t$, we  use for the  velocity along the wall normal  
\begin{equation}
v_W=\cos\theta + \frac{3p(3\cos^2\theta-1)}{8h^2},
\label{Eq:pvw}
\end{equation}
where we  included $v_{\text{HI}}$ for a force-dipole swimmer \cite{Spagnolie12}.
Since at the wall $v_W<0$ for the pusher  [Eq.~(\ref{Eq:pvw})], $T_t>T_t^{\text{ABP}}$. 
So, the region-I pusher 
reaches an even smaller height than the ABP
($\Delta h < \Delta h^{\text{ABP}}$).

In region II (large $\text{Pe}_r$ and large $p$) the mean detention time is $T \sim \exp(3\text{Pe}_r p \cos^2\theta^{*}/8)$ (see Eq.~(6) of main text) with $\cos\theta^{*} = (-4+\sqrt{16+27p^2})/(9p)$.
On the other hand, 
for distances up to $\Delta h$  we have the lower limit $|v_{\text{HI}}| > 3p/[8(\Delta h +1)^2]=\text{const}$  [Eq.~(\ref{Eq:pvw})] and we use 
 Eq.~(\ref{Eq:Tt}) to obtain a lower bound for $T_t$.
In the limit $\text{Pe} \rightarrow \infty$ [which applies in region II due to Eq.~(\ref{Eq:PePer})], the pusher needs the time  $T_t \gtrsim \exp[(3/4) p\text{Pe}\Delta h/(\Delta h +1)^2]$ [see Eq.~(\ref{Eq:Tt2})]
to reach a distance $\Delta h$ from the wall with the stable orientation $\theta_s=\pi/2$.
Hence, by comparing $T_t$ with $T$,
we obtain $\Delta h/(\Delta h +1)^2 < (-4+\sqrt{16+27p^2})^2/(486p^2)$, and $\Delta h < 0.063$ for all $p$.

So, for pushers in region I and region II  the escape process is always determined by rotational and not translational motion.
However, fluctuations in $h$ may alter the detention times due to the $h$-dependent hydrodynamic swimmer-wall interactions,
as shown in Sec.~3.4.


\vspace{4mm}\noindent\textsl{3.3.3~Puller} \\[-2mm]

A puller ($p<0$) with sufficiently large $|p|$ and $\text{Pe}_r$ has a stable orientation $\theta_s \approx \pi$ at the wall, and its mean detention time is
$T \sim \exp(3|p|\text{Pe}_r/8)$ (Eq.~(7) of the main text).
By using Eq.~(\ref{Eq:Tt2}) and Eq.~(\ref{Eq:pvw}),    the time to reach a height $\Delta h$  via translational diffusion 
is 
$T_t \gtrsim \exp[2\text{Pe}\Delta h (1+ (3/4)|p|/(\Delta h +1 )^2)]$
since  $|v_{\text{HI}}| \ge 1 +(3/4)|p|/(\Delta h+1)^2$ [see Eq.~(\ref{Eq:pvw})].
Comparing now $T_t$ and $T$, and using Eq.~(\ref{Eq:PePer}),  we obtain $(3/8)|p| < 6\Delta h (1+(3|p|)/(4(\Delta h + 1)^2))$, which results in $\Delta h < 0.1$.

For a puller with small $|p|\text{Pe}_r$ the detention time $T \approx T^{\text{ABP}} < 2\text{Pe}_r$.
Since $v_W<0$ around the stable orientation, $T_t>T_t^{ABP}$ and $\Delta h <\sqrt{2/3}$.

So, also a puller
is not able to diffuse far from the wall 
although, similar to the pusher,
fluctuations in $h$ may influence the detention times as shown in Sec.~3.4.


\vspace{4mm}\noindent\textsl{3.3.4~Source Dipole Swimmer} \\[-2mm]

The source dipole swimmer reorients faster than the ABP at the wall towards the escape angle ($T<T^{\text{ABP}}$).
The velocity along the wall normal reads $v_W=\cos\theta- q \cos\theta/h^3<0$ \cite{Spagnolie12} for orientations towards the wall 
such that $T_t>\text{Pe}\Delta h^2$ [Eqs.~(\ref{Eq:Tt})] for any P\'eclet number,
and $\Delta h <\sqrt{2/3}$. So, also the source dipole swimmer does not leave the wall via translational diffusion.


\vspace{5mm}\noindent\textsl{3.4~Comparison of 1D and 2D Model} \\

\begin{figure}[bt]
\includegraphics[width=0.95\columnwidth]{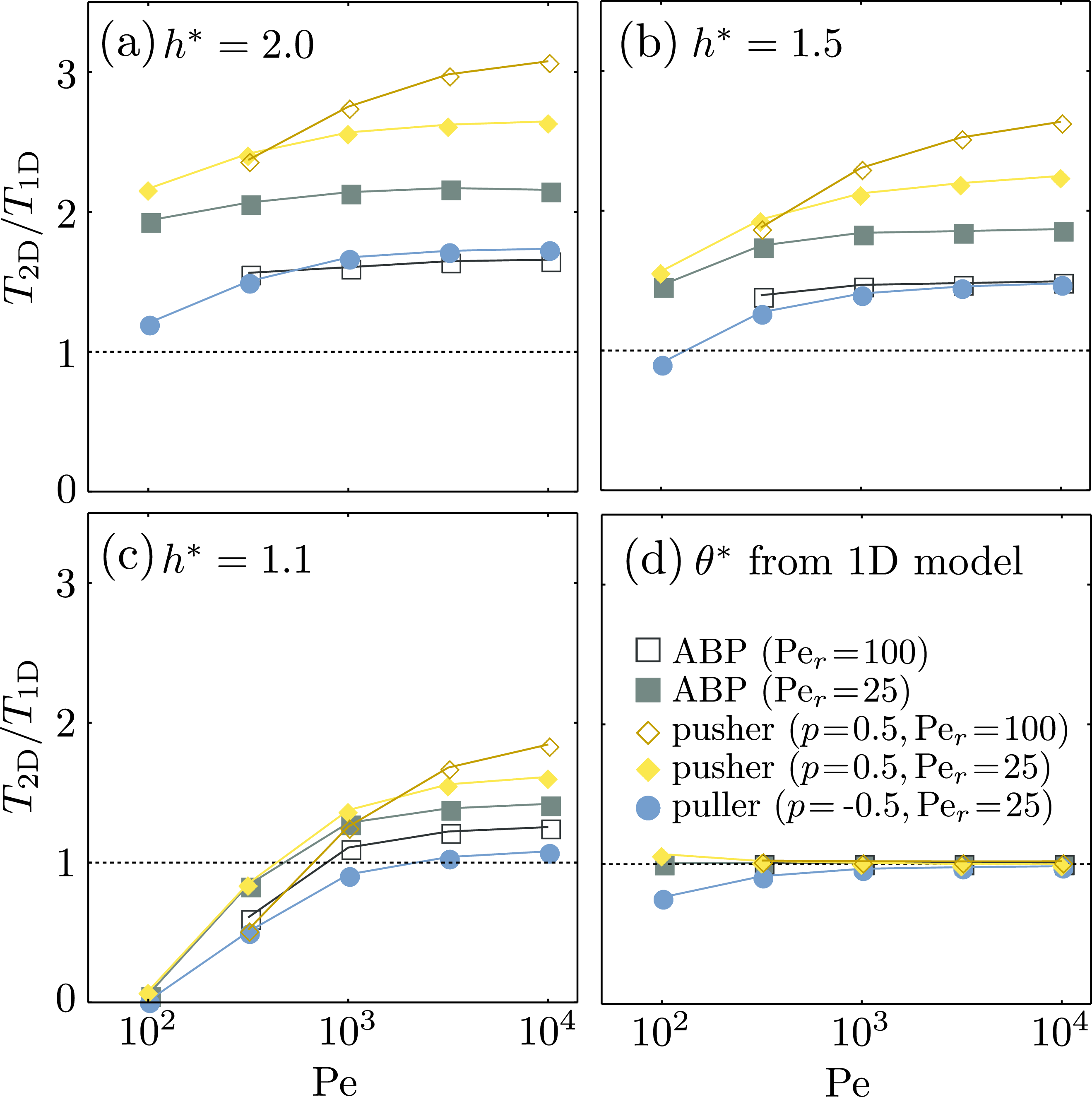}
\caption{(a-c) Comparison of 2D model with 1D model for the incoming angle $\theta_0=3\pi/4$ for different escape heights $h^{*}$. In (d)
the mean detention time for reaching the escape angle $\theta^{*}$ in the full 2D model
is compared to the 1D model.}
\label{Fig:1D2D}
\end{figure}

In the main text we introduced  a 1D model [Eq.~(2) of main text] 
for calculating distributions of the  swimmer-wall detention times but also outlined a 2D model [Eq.~(9) of main text].
While in the 1D model the escape from the surface is defined by reaching an escape angle $\theta^{*}$, in the 2D model the swimmer  escapes 
 when it reaches a specific height $h^{*}$ above the wall.
In Fig.~\ref{Fig:1D2D}(a-c) we now compare the mean detention times determined from both models for different swimmer types, persistence numbers, P\'eclet numbers, and escape distances $h^{*}=1 + \Delta h$
for the incoming angle $\theta_0=3\pi/4$.
We observe that for small heights and sufficiently small Pe, $T_{\text{2D}}<T_{\text{1D}}$ since during reorientation at the wall the swimmers can reach  $h^{*}$  via translational
diffusion. For sufficiently large  $h^{*}$,  one always finds  $T_{\text{2D}}>T_{\text{1D}}$,   since the swimmer needs additional time to reach  $h^{*}$  after leaving the surface.

Reorientation rates due to hydrodynamic swimmer-wall interactions, $\Omega_{\text{HI}}$,  depend in general on the distance from the wall \cite{Spagnolie12}.
Hence, fluctuations in the position $h$ above the surface may influence the detention times
for reaching the escape angle $\theta^{*}$  defined in the 1D model.
In Fig.~\ref{Fig:1D2D}(d) we show the mean detention time $T_{\text{2D}}$ by solving the full Langevin dynamics for an incoming angle $\theta_0$  and compare
 it to $T_{\text{1D}}$ from the 1D model.
Depending on the specific swimmer type, translational noise can either enhance or decrease the detention times for sufficiently small Pe.
However, the effect of  translational noise is in general rather small.
As expected, for active Brownian particles with $\Omega_{\text{HI}}=0$ the detention times are independent of translational noise.


\vspace{5mm}\noindent\textbf{4.~Effect of Steric Interactions for "Chlamydomonas"} \\

\begin{figure}
\includegraphics[width=0.7\columnwidth]{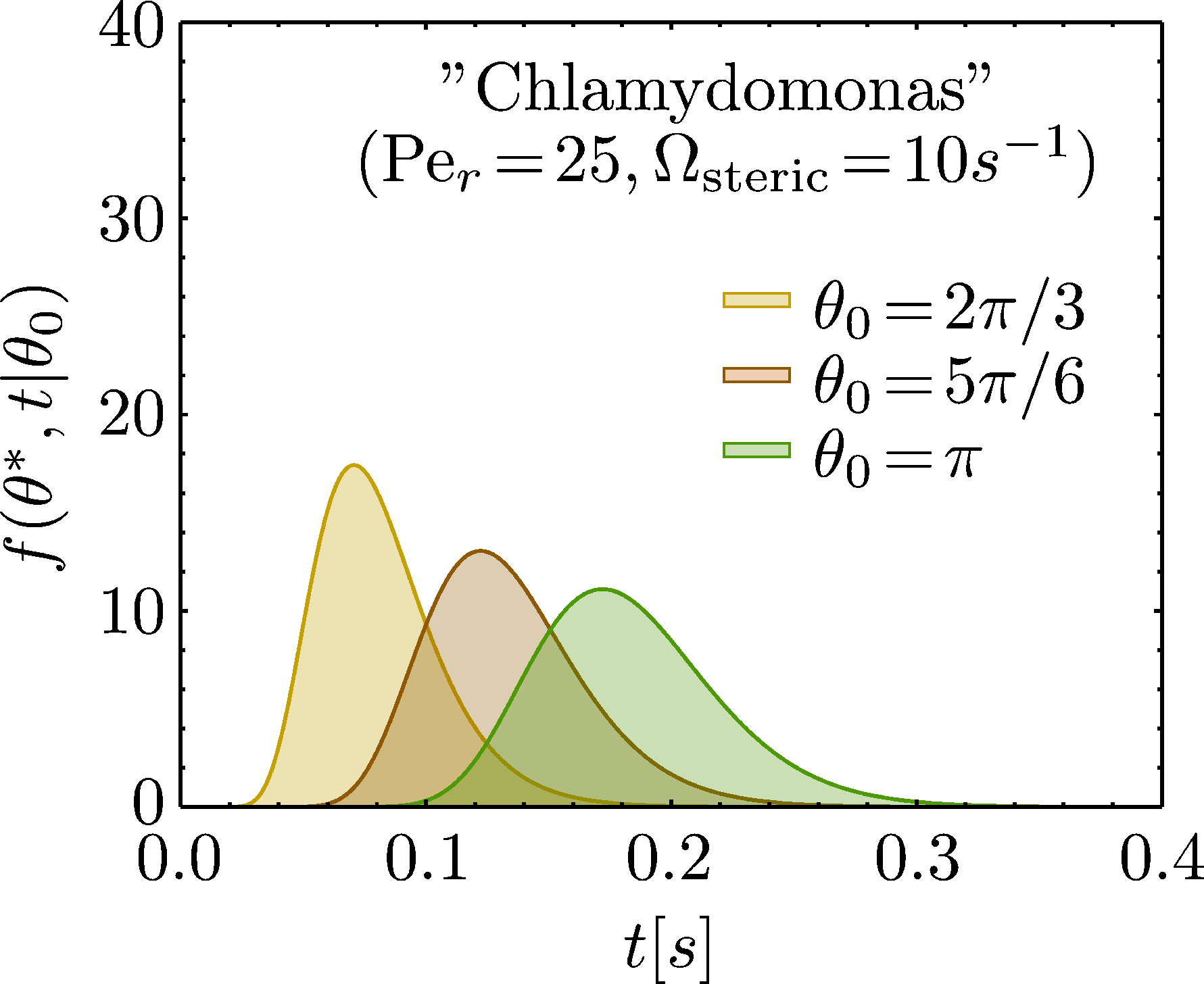} 
\caption{Detention time distributions $f(\theta^\ast,t|\theta_0)$ for a simple Chlamydomonas model ($\text{Pe}_r=25$, $\Omega_\text{steric}=10s^{-1}$, $\theta^{*}=0.64\pi$) for different incoming angles $\theta_0$.}
\label{Fig:Chlamy}
\end{figure}

Our model is  able to treat  steric swimmer-wall interactions, which can  occur for any swimmer with flagella-wall contact.
As an example we show DTDs of a simplified model for \textit{Chlamydomonas}.
We neglect the hydrodynamic swimmer-wall interactions  
and simplify the time-dependent flagella-wall interactions by assuming a constant reorientation rate $\Omega_{\text{steric}}>0$,
which rotates the cell away from the wall  as discussed in \cite{Kantsler13}.
In our 1D model we use
 realistic values for the persistence number ($\text{Pe}_r=25$),  a reorientation rate ($\Omega_{\text{steric}}=10\text{rad}/s$),
and an escape angle ($\theta^{*}=0.64\pi$) as suggested in Ref.~\cite{Kantsler13}.
Typical DTDs for different incoming angles $\theta_0$ are shown in Fig.~\ref{Fig:Chlamy}.
In accordance with Ref.~\cite{Kantsler13},
the detention time of a Chlamydomonas swimmer at the wall is always less than half a second.


\begin{figure}
\includegraphics[width=0.6\columnwidth]{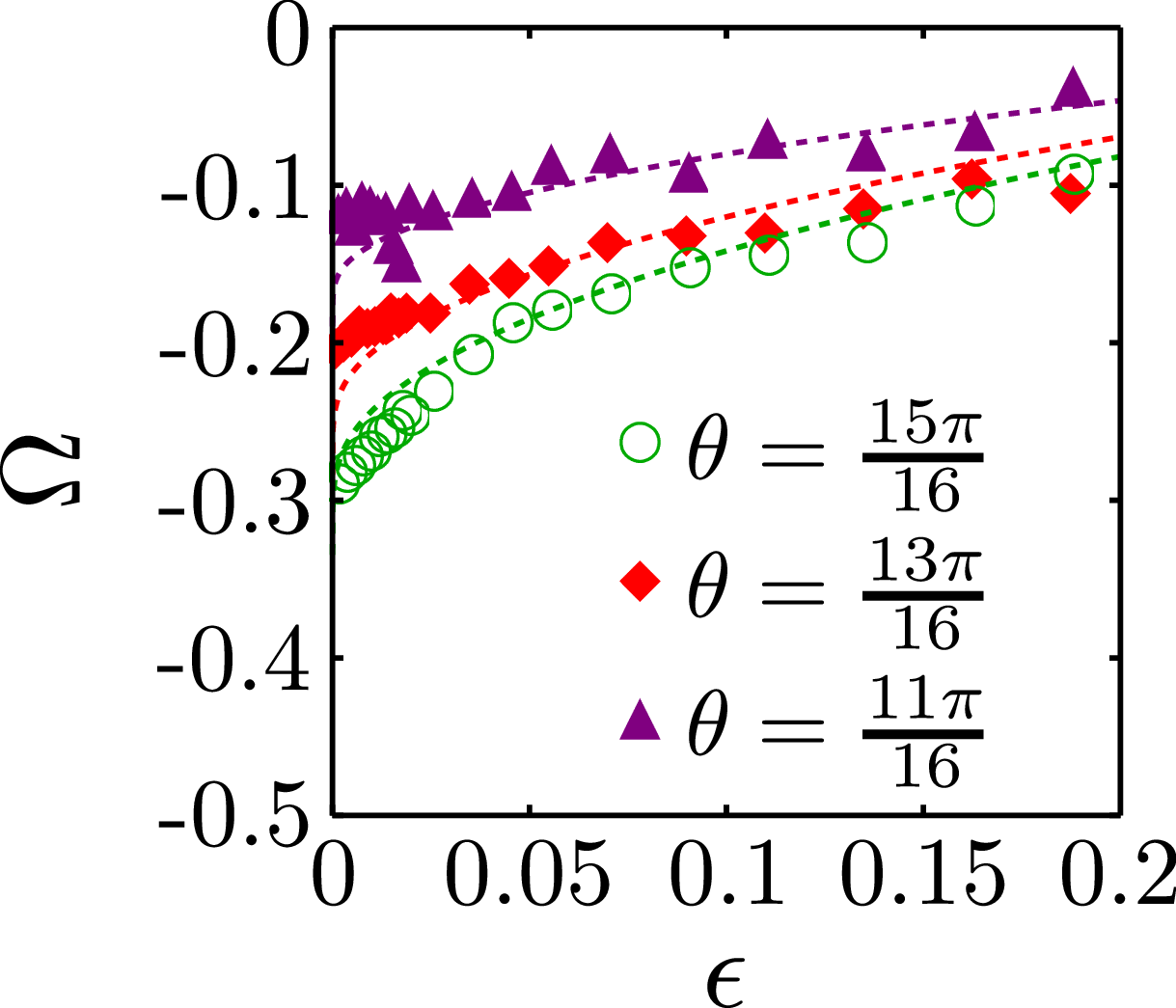}
\caption{Distance-dependent angular velocity of a neutral squirmer in front of a no-slip wall for several orientations $\theta$
compared to lubrication theory (dashed lines).}
\label{Fig:lubri}
\end{figure}


\vspace{5mm}\noindent\textbf{5.~MPCD Simulations} \\

We perform MPCD simulations by using the same parameters to model the fluid and the squirmer as in \cite{Zoettl14}.
To determine the constant $c$ of Eq.~(8) 
in
the main text, we measure the wall-induced angular velocity $\Omega(\epsilon,\theta)$ of 
a large number of neutral squirmers in front of a no-slip wall in separate simulations  to average out thermal noise.
Figure~\ref{Fig:lubri} shows $\Omega(\epsilon)$ for different $\theta$ compared to the analytic expression obtained by using Eq.~(8) of the main text and $\Omega=-M/\gamma_r$.
The curves fit best for $c \approx 0.9$.

We then perform  simulations of squirmer-wall interactions for different incoming angles $\theta_0$ and measure the detention times at the wall. 
The mean distance from the wall averaged over all trajectories is $\epsilon \approx 0.01$.


\begin{thebibliography}{10}
\bibitem{noteTrans}
The escape from a surface is indeed mainly determined by rotational motion, see \cite{supp}.
\bibitem{timedependent}
We note that swimmer flow fields and hence $\Omega_{\text{HI}}$ can be time-dependent  \cite{DrescherGuasto10} and may influence the DTDs, which we neglect in the following.
\bibitem{fn2}
We neglect here the small translational-rotational coupling.
\end{thebibliography}
\end{document}